\documentclass[12pt]{article}
\usepackage{amsmath}
\usepackage{graphicx,psfrag,epsf}
\usepackage{enumerate}
\usepackage{natbib}
\usepackage[colorlinks=true,linkcolor=black,citecolor=black,urlcolor=black]{hyperref}

\newcommand{\blind}{0}

\usepackage{dsfont}    

\providecommand{\realline}{\mathds{R}}
\providecommand{\prob}{\mathrm{P}}
\providecommand{\differential}{\mathrm{d}}
\providecommand{\expect}{\mathrm{E}}
\providecommand{\relEnt}{\mathcal{D}_\mathrm{KL}}
\providecommand{\symKL}{\mathcal{D}_\mathrm{s}}

\begin{document}

\def\spacingset#1{\renewcommand{\baselinestretch}%
{#1}\small\normalsize} \spacingset{1}


\if0\blind
{
  \title{\bf Discrete approximation\\ of a mixture distribution\\ via restricted divergence}
  \author{Christian R\"over
    and 
    Tim Friede\thanks{This project has received funding 
    from the European Union's \textsl{Seventh
    Framework Programme for research, technological development and
    demonstration} under grant agreement number FP~HEALTH~2013-602144
    ``\textsl{Innovative methodology for small populations research (InSPiRe)}''.}\hspace{.2cm}\\[1ex]
    Department of Medical Statistics, \\University Medical Center G\"{o}ttingen}
  \maketitle
} \fi

\if1\blind
{
  \bigskip
  \bigskip
  \bigskip
  \begin{center}
    {\LARGE\bf Title}
\end{center}
  \medskip
} \fi

\bigskip
\begin{abstract}
  Mixture distributions arise in many application areas, for example
  as marginal distributions or convolutions of distributions.
  We present a method of constructing an easily tractable discrete
  mixture distribution as an approximation to a mixture distribution
  with a large to infinite number, discrete or continuous, of
  components.  
  The proposed \textsc{direct} (Divergence Restricting Conditional Tesselation)
  algorithm is set up such that a pre-specified precision, defined in
  terms of Kullback-Leibler divergence between true distribution and
  approximation, is guaranteed.
  Application of the algorithm is demonstrated in two examples.
\end{abstract}

\noindent%
\vfill

\noindent\textit{Reference as:}\\
C.~R\"{o}ver, T.~Friede (2017). Discrete approximation of a mixture distribution via restricted divergence. \textsl{Journal of Computational and Graphical Statistics} \textbf{26}(1), 217--222.\\
\href{https://doi.org/10.1080/10618600.2016.1276840}{doi:$\,$10.1080/10618600.2016.1276840}
\vspace{2cm}

\newpage
\spacingset{1.45} 
\section{Introduction}
  Mixture distributions with a large to infinite number of mixture
  components commonly occur in many fields of application
  \citep[e.g.,][]{Seidel2010}.  Common examples include e.g.\ marginal
  (posterior) distributions, convolutions of random variables,
  predictive distributions, distributions of test statistics,
  overdispersed sampling distributions, and many more.

  If the mixture distribution's exact marginal density, distribution
  or quantile functions are not available in analytical form, then
  practical application of such mixtures is often very limited.  Such
  mixtures may then often be approximated to a sufficient degree by a
  mixture of a lower, \textsl{finite} number of components.  How
  exactly to select such a finite set of components however is not
  obvious. In the following we describe a general approach and an
  algorithm allowing to set up a finite mixture as an approximation to
  a mixture distribution with a large or infinite number of components
  in a completely automated way.  The construction is based on the
  \textsl{Kullback-Leibler divergence} or \textsl{relative entropy}
  between distributions and as such aims at bounding the (expected)
  logarithmic ratio of exact and approximate probability densities.

\section{Kullback-Leibler divergence}\label{sec:KLDiv}
\subsection{Definitions}
  The \textsl{Kullback-Leibler divergence} or \textsl{relative
    entropy} of two probability distributions with probability density
  functions $p$ and $q$ is defined as the \textsl{expected logarithmic
    ratio of densities} with respect to the former distribution ($p$),
  \begin{equation}
    \relEnt\bigl(p(\theta)\big\|q(\theta)\bigr) 
    \;=\; \int_\Theta \log\Bigl(\frac{p(\theta)}{q(\theta)}\Bigr)\,p(\theta)\,\differential \theta
    \;=\; \expect_{p(\theta)}\biggl[ \log\Bigl(\frac{p(\theta)}{q(\theta)}\Bigr)\biggr]
  \end{equation}
  \citep[Ch.~2]{InfoTheory}.
  In case of discrete probability distributions $p$ and $q$, the
  integrals simplify to sums, but for simplicity we will stick to the
  integral notation in the following.  The relative entropy is always
  positive, it is zero if the two distributions are identical ($p=q$),
  and larger otherwise.  The divergence (in general) is \textsl{not}
  symmetric:
  $\relEnt\bigl(p(\theta)\big\|q(\theta)\bigr) 
   \;\neq\; 
   \relEnt\bigl(q(\theta)\big\|p(\theta)\bigr)$.
  The \textsl{symmetrized (KL-) divergence} is defined as
  \begin{equation}
    \symKL\bigl(p(\theta)\big\|q(\theta)\bigr) 
    \;=\;
    \relEnt\bigl(p(\theta)\big\|q(\theta)\bigr) 
    + \relEnt\bigl(q(\theta)\big\|p(\theta)\bigr)
  \end{equation}
  \citep{KullbackLeibler1951}. 
  Unlike the \textsl{directed} divergence, $\symKL$ is obviously symmetric.
  Note that, trivially but importantly, 
  \begin{equation}
    \symKL\bigl(p(\theta)\big\|q(\theta)\bigr) 
    \;\geq\;
    \max\bigl\{\relEnt\bigl(p(\theta)\big\|q(\theta)\bigr),\, 
               \relEnt\bigl(q(\theta)\big\|p(\theta)\bigr)\bigr\}\mbox{,}
  \end{equation}
  i.e., the symmetrized divergence bounds both individual directed
  divergences.  For simplicity, in the following we will mostly be
  focusing on \textsl{symmetrized} KL-divergences.

  For example, the Kullback-Leibler divergence for two normal
  distributions with mean and variance parameters $(\mu_A,\sigma^2_A)$
  and $(\mu_B,\sigma^2_B)$, respectively, is given by
  \begin{eqnarray}
    \relEnt\bigl(p(\theta|\mu_A,\sigma_A)\big\|p(\theta|\mu_B,\sigma_B)\bigr) 
    &=& \textstyle
    \frac{1}{2} \Bigl( \frac{(\mu_A-\mu_B)^2}{\sigma_B^2}
                       + \frac{\sigma_A^2}{\sigma_B^2}
                       + \log\bigl(\frac{\sigma_B^2}{\sigma_A^2}\bigr) 
                       - 1 \Bigr)
  \end{eqnarray}
  \citep[Ch.~9]{Kullback1959}.
  The symmetrized divergence then results as
  \begin{eqnarray}\label{eqn:normalDiv}
    \symKL\bigl(p(\theta|\mu_A,\sigma_A)\big\|p(\theta|\mu_B,\sigma_B)\bigr) 
    &=& \textstyle
    \frac{(\mu_A-\mu_B)^2}{\left(\frac{1}{2}(\sigma_A^{-2}+\sigma_B^{-2})\right)^{-1}}
    + \frac{(\sigma_A^2-\sigma_B^2)^2}{2\,\sigma_A^2\,\sigma_B^2}\mbox{.}
  \end{eqnarray}

\subsection{Motivation and interpretation}
  The Kullback-Leibler divergence is generally regarded as a measure
  of \textsl{discrepancy} between probability distributions.  For
  example, when a simple parametric approximation to a more
  complicated distribution is sought, the approximation may reasonably
  be matched against the true distribution via minimization of the
  divergence \citep{BernardoSmith,OHagan}.

  The divergence~$\relEnt$ relates to the logarithmic ratio of
  densities.  The domain of main interest here is the limit of very
  similar $p$ and $q$, i.e., almost equal numerator and denominator,
  when the density ratio is close to unity.  In that case the
  logarithmic ratio approximately corresponds to the ``relative
  difference'' in densities: since $\log(x) \approx x-1$ for $x
  \approx 1$ (and hence $\log(a/b) \approx a/b-1$ for $a \approx b$),
  a divergence of, say, 0.01 approximately corresponds to an
  (expected) 1\% difference between numerator and denominator.

  While there is no simple connection relating the divergence of two
  distributions to their moments, one can get an impression by
  considering the generic case of two normal distributions.
  For some fairly obvious parameter choices we get:
  \begin{eqnarray}
    \sigma_B\!=\!\sigma_A,\;\;
    \mu_B\!=\!\mu_A+c\sigma_A
    &\qquad\Rightarrow\qquad&
    \relEnt(p \| q) = {\textstyle\frac{1}{2}} c^2\mbox{,} \\
    &&\symKL(p \| q) = c^2\mbox{,}
  \end{eqnarray}
  and
  \begin{eqnarray}
    \mu_B\!=\!\mu_A,\;\;
    \sigma_B \!=\! (1\!+\!c)\sigma_A
    &\qquad\Rightarrow\qquad&
    \relEnt(p \| q) = {\textstyle\frac{1}{2(1+c)^2} + \log(1\!+\!c) - \frac{1}{2}}
                         \approx c^2\mbox{,}\\
    && \symKL(p \| q) = {\textstyle\frac{c^2 (c+2)^2}{2(c+1)^2}}
                           \approx 2c^2\mbox{,}
  \end{eqnarray}
  where the latter approximations follow from Taylor expansion 
  around $c=0$.

  From the above we can see that, for example, for equal variances, a
  difference in means by, say, $1\%$ of a standard deviation
  corresponds to a symmetrized divergence $\symKL=0.01^2=0.0001$. For
  equal means on the other hand, standard deviations differing by
  $1\%$ correspond to a symmetrized divergence of $\approx 0.0002$.

\section{Mixture distributions and discrete approximations}\label{sec:mixture}
\subsection{Definitions}
  Suppose a random variable $Y$ follows a distribution with density
  $p(y|x)$ that depends on a parameter~$x$. If that parameter is not
  fixed, but again is a random variable ($X$) with density $p(x)$,
  then the (marginal) distribution of $Y$ is called a \textsl{mixture
    distribution}.  The joint density of $X$ and $Y$ is given by
  $p(x,y)= p(y|x)\times p(x)$. What is commonly of interest is the
  marginal (unconditional) distribution of $Y$, whose density results
  by integration as $p(y)=\int p(x,y) \,\differential x = \int p(y|x)
  \,p(x)\,\differential x$. The (marginal) distribution of the
  underlying variable that is conditioned upon, $p(x)$, is called the
  \textsl{mixing distribution} \citep{Seidel2010}
  or \textsl{latent distribution} \citep{Lindsay}.  

  Mixture distributions arise frequently in statistical problems, for
  example as marginal (posterior) distributions or as convolutions of
  random variables.  In the following we will assume that $X$ is
  one-dimensional, and that the domain of~$X$ is the real line, or a
  subset thereof (continuous or discrete).

\subsection{Binning}
  In order to transition from continuous to discrete mixtures, we
  define a \textsl{binning} of the domain of $X$.  Let $\{x_{(1)},
  x_{(2)}, \ldots, x_{(k-1)}\}\subset \realline$ be a set of bin
  margins with $x_{(1)} < x_{(2)} < \cdots < x_{(k-1)}$. These define
  the (exhaustive and disjoint) set of $k$~bins
  $\{\mathcal{X}_i\}_{i=1,\ldots,k}$ with
  \begin{equation}
    \mathcal{X}_i \;=\; 
    \left\{ \small \begin{array}{ll} 
              \{x: x \leq x_{(1)}\} & \mbox{if } i=1 \\
              \{x: x_{(i-1)} < x \leq x_{(i)}\} & \mbox{if } 1<i<k \\
              \{x: x_{(k-1)} < x\} & \mbox{if } i=k\mbox{.}
            \end{array}
    \right. 
  \end{equation}
  In addition, the set of $k$~points
  $\{\tilde{x}_1,\ldots,\tilde{x}_k\}$ with $\tilde{x}_i\in
  \mathcal{X}_i$ defines a set of \textsl{reference points}, one for
  each bin.  Each bin also has a probability~$\pi_i$ (with respect
  to~$p(x)$) associated, which is given by
  \begin{equation}
    \pi_i \;=\; \prob\bigl(x_{(i-1)} < x\leq x_{(i)}\bigr)
          \;=\; \prob\bigl(x\in \mathcal{X}_i\bigr)\mbox{.}
  \end{equation}

\subsection{The binned mixture}
  In addition to the probability density~$p(x,y)$ given above, we
  define another probability distribution with density~$q(x,y)$ that
  has the same marginal density (mixing distribution)
  \begin{equation}
    q(x) \;=\; p(x)\mbox{,}
  \end{equation}
  and whose conditional probability density is given by
  \begin{equation}
    q(y|x) \;=\; 
    p(y|x\!=\!\tilde{x}_i) \quad \mbox{for } x \in \mathcal{X}_i\mbox{.}
  \end{equation}
  So $q$ is similar to $p$, but instead of conditioning on the
  ``exact'' $x$ value as in the original definition above, this
  probability distribution conditions on the corresponding bin's
  reference value $\tilde{x}_i$, depending on which bin~$x$ belongs
  to.  The joint distribution of $X$ and $Y$ again is defined through
  its joint density: $q(x,y) = q(x) \times q(y|x)$. The marginal
  density of $Y$ again turns out as $q(y) = \int q(y|x)
  \,q(x)\,\differential x$.  Equivalently, the binning may be
  considered a discretization of the mixing distribution while keeping
  the conditional distribution the same. The discretized mixing
  distribution simply has the reference points
  $\{\tilde{x}_1,\ldots,\tilde{x}_k\}$ as its domain, while the
  associated bin probabilities $\{\pi_1,\ldots,\pi_k\}$ define the
  probability mass function. The reference points consequently act as
  ``support points'' for the discretized mixing distribution here;
  alternating between these points of view is sometimes helpful.

  This ``binned'' approximation to the joint distribution of $(X,Y)$
  is useful, as the resulting marginal distribution of $Y$, $q(y)$, is
  a discrete sum of conditional densities (rather than an integral),
  making numerical evaluation very easy. The marginal density
  simplifies to
  \begin{equation}\label{eqn:mixtureApprox}
    q(y) \;=\; \sum_{i=1}^k \pi_i \, p(y|\tilde{x}_i)\mbox{.}
  \end{equation}
  Analogously, the cumulative distribution function (CDF) may also be
  expressed as a weighted sum of the component CDFs.  Random number
  generation as well as computation of moments for finite mixtures is
  also straightforward \citep{Lindsay}.

\section{Constructing binned mixture approximations}\label{sec:construct}
\subsection{Some preliminary results}
  For each bin~$i$ define the maximum symmetrized KL-divergence
  \begin{equation}
    d_i \;=\;
    \max_{x \in \mathcal{X}_i} \Bigl\{ \symKL \bigl( p(y|x) \big\| p(y|\tilde{x}_i)\bigr) \Bigr\}
    \;=\;
    \max_{x \in \mathcal{X}_i} \Bigl\{ \symKL \bigl( p(y|x) \big\| q(y|x)\bigr) \Bigr\}\mbox{,}
  \end{equation}
  i.e., the maximum (symmetrized) divergence between distributions
  $p(y|x)$ corresponding to points within the $i$th bin and the
  corresponding $i$th reference point.

  The \textsl{chain rule for relative entropy} states that
  \begin{equation}
    \relEnt\bigl(p(x,y) \big\| q(x,y)\bigr)
    \; = \;
    \relEnt\bigl(p(x) \big\| q(x)\bigr) 
    + \expect_{p(x)}\Bigl[\relEnt\bigl(p(y|x) \big\| q(y|x)\bigr)\Bigr] \label{eqn:condRelEnt}
  \end{equation}
  \citep[Sec.~2.5]{InfoTheory}.  
  In other words, the divergence of two joint distributions is the sum
  of the divergence of the mar\-gin\-als and the expected divergence
  of the conditionals.  Note that the expectation
  in~(\ref{eqn:condRelEnt}) is also known as the \textsl{conditional
    relative entropy} \citep[Sec.~2.5]{InfoTheory}.  For the
  \textsl{symmetrised} divergence immediately follows an analogous
  property:
  \begin{equation}
    \symKL\bigl(p(x,y) \big\| q(x,y)\bigr) \\
    \;=\;
    \symKL\bigl(p(x) \big\| q(x)\bigr) 
    + 
    \expect_{p(x)}\Bigl[\symKL\bigl(p(y|x) \big\| q(y|x)\bigr)\Bigr]\mbox{.} \label{eqn:SymmetrizedChainRule}
  \end{equation}

  In our case we have identical marginal distributions for~$X$ under
  both distributions, $p(x)\!=\!q(x)$, so that
  \begin{equation}
    \relEnt\bigl(p(x) \big\| q(x)\bigr) \;=\; \relEnt\bigl(q(x) \big\| p(x)\bigr) \;=\; 0
  \end{equation}
  and consequently
  \begin{eqnarray}
    \symKL\bigl(p(x,y) \big\| q(x,y)\bigr) 
    &=& 
    \expect_{p(x)}\Bigl[\symKL\bigl(p(y|x) \big\| q(y|x)\bigr)\Bigr]\mbox{.} \label{eqn:margKL}
  \end{eqnarray}

  We are interested in the approximation of $p(x,y)$ through the
  simplified distribution $q(x,y)$, and in particular of $p(y)$ by
  $q(y)$.  We know, again via the chain rule, that
  \begin{eqnarray}
    \label{eqn:boundDiv}
    \symKL\bigl(p(y) \big\| q(y)\bigr)
    &\stackrel{\mbox{\footnotesize(\ref{eqn:SymmetrizedChainRule})}}{=}& 
    \symKL\bigl(p(x,y) \big\| q(x,y)\bigr)
    - \expect_{p(x)}\Bigl[\symKL\bigl(p(x|y) \big\| q(x|y)\bigr)\Bigr]
    \\
    &\leq& 
    \symKL\bigl(p(x,y) \big\| q(x,y)\bigr)
    \\
    &\stackrel{\mbox{\footnotesize(\ref{eqn:margKL})}}{=}& 
    \expect_{p(x)}\Bigl[\symKL\bigl(p(y|x) \big\| q(y|x)\bigr)\Bigr]
    \\
    \label{eqn:boundSum}
    &\leq& 
    \sum_i \pi_i d_i
    \\
    \label{eqn:boundMax}
    &\leq& 
    \max_i d_i\;=:\;\delta\mbox{.}
  \end{eqnarray}
  So, by limiting the divergences of conditionals $p(y|x)$ and
  $q(y|x)$ within each single bin such that these remain~$\leq\delta$
  (\ref{eqn:boundMax}), we can now also bound the divergence of exact
  and approximate marginals $p(y)$ and $q(y)$ (\ref{eqn:boundDiv}).

\subsection{The proposed approach}
  Given the bin-wise divergences, we can now bound the divergence of
  exact and discretized marginals. The obvious question now is whether
  and how one can invert the argument and construct a grid
  approximation matching a pre-specified maximum divergence~$\delta$.
  For a given (reference) point~$x$ in the mixing distribution's
  domain, we can find a corresponding neighbourhood within which the
  divergence remains below~$\delta$. Once we have defined a single bin
  this way, we can also generate an exhaustive covering of the whole
  parameter space through such bins.
  We abbreviate this method as the \textsl{\textsc{direct} (Divergence
    Restricting Conditional Tesselation) approach}, as it aims at a
  covering of the conditional's parameter space while bounding the
  divergence.

  In some cases it is not possible to have a finite number of bins
  associated with finite bin-wise divergences.
  A ``trick'', if necessary, then is to simply ignore some fraction of
  parameter space (of the mixing distribution's domain) that is
  associated with a pre-set, arbitrarily small probability~$\epsilon$
  and do the binning on the remaining share of parameter space.
  Problems with unbounded divergences, or infinite numbers of
  necessary bins, commonly occur towards one or both of the parameter
  space's margins.  Neglecting a certain fraction of parameter space
  that is associated with an (arbitrarily) small
  probability~$\epsilon$ will usually not pose a significant practical
  problem, as it will only add another bit to the error budget that
  needs to be considered in (almost) any numerical computation anyway.

\subsection{The sequential {\sc direct} algorithm}\label{sec:algorithm}
  We will in the following construct a binning so that the resulting
  discrete approximation of the exact marginal does not differ, in
  terms of symmetrized divergence, and with that of both directed
  divergences, from the exact (``continuous'') marginal by more than a
  pre-specified amount. The number ($k$) of components and the
  placement of reference points will be determined automatically in
  the process.  The idea is to sequentially divide the mixing
  distribution's domain into bins, while firstly ensuring that the
  divergences within bins are bounded, and secondly, if necessary,
  ignoring the mixing distribution's extreme left and/or right tails.
  In order to proceed, in the following we will assume that the
  divergence between any pair of points ($x_1, x_2$) in parameter
  space is Lipschitz continuous, at least with\-in a range
  $[\tilde{x}_1, \tilde{x}_k]$ with $\prob \bigl(X \notin
  [\tilde{x}_1, \tilde{x}_k]\bigr) \leq \epsilon$.  This will ensure
  that the algorithm will work, although violations do not necessarily
  prevent a solution; even continuity is not strictly necessary.
  A possible implementation of the \textsc{direct} approach 
  is defined in Tab.~\ref{tab:direct}.

  \begin{table}[t]
  \caption{\label{tab:direct}The sequential \textsc{direct} algorithm (see Sec.~\ref{sec:algorithm}).}
  \begin{center} \sffamily
  \begin{tabular}{rl}
    \hline \hline
    1. & \begin{minipage}[t]{0.8\linewidth}
         Specify a maximum KL-divergence $\delta>0$, 
         some small probability $0 \leq \epsilon \ll 1$,
         and a starting reference point $\tilde{x}_1$.
         Sensible values for $\tilde{x}_1$ may for example 
         be the minimum possible value, the $\frac{\epsilon}{2}$-quantile,
         or any value with $\prob(X\leq\tilde{x}_1)<\epsilon$.
         Define $\epsilon_1 := \prob(X\leq\tilde{x}_1)\geq 0$.
         Set $i=1$.
         \end{minipage}\\
    2. & \begin{minipage}[t]{0.8\linewidth}
         Set $x^\star=\tilde{x}_1$. 
         Obviously, $\symKL \bigl( p(y|\tilde{x}_1) \big\| p(y|x^\star)\bigr)=0$.
         Now increase $x^\star$ as far as possible while ensuring that
         $\symKL \bigl( p(y|\tilde{x}_1) \big\| p(y|x^\star)\bigr)\leq\delta$. 
         Use this point as the first bin margin: $x_{(1)}=x^\star$. 
         Compute $\pi_1=\prob(x<x_{(1)})$. Set $i=i+1$.
         \end{minipage}\\
    3. & \begin{minipage}[t]{0.8\linewidth}
         Increase $x^\star$
         until $\symKL \bigl( p(y|x_{(i-1)}) \big\| p(y|x^\star)\bigr)=\delta$.
         Use this point as the next reference point: $\tilde{x}_{i} = x^\star$. 
         \end{minipage}\\
    4. & \begin{minipage}[t]{0.8\linewidth}
         Increase $x^\star$ again
         until $\symKL \bigl( p(y|\tilde{x}_{i}) \big\| p(y|x^\star)\bigr)=\delta$.
         Use this point as the next bin margin: $x_{(i)}=x^\star$.
         \end{minipage}\\
    5. & \begin{minipage}[t]{0.8\linewidth}
         Compute the bin weight $\pi_i=\prob(x_{(i-1)} < X \leq x_{(i)})$.
         \end{minipage}\\
    6. & \begin{minipage}[t]{0.8\linewidth}
         If $\prob(X > x_{(i)}) > (\epsilon-\epsilon_1)$, set $i=i+1$ and proceed at step~3.
         Otherwise stop.
         \end{minipage}\\
    \hline \hline
  \end{tabular}
  \end{center}
  \end{table}

  Reference points~$\tilde{x}_i$ and corresponding weights~$\pi_i$ now
  allow to define an approximation~$q$ as in
  (\ref{eqn:mixtureApprox}).  It is actually not necessary to also
  keep track of the exact bin margins~$x_{(i)}$ once the bin
  weights~$\pi_i$ are determined.  The maximum divergence of
  conditionals, and with that of the marginals, will now be~$=\delta$,
  possibly up to a bit of probability ($\leq\epsilon$) beyond the
  first and/or last bins.

  The essence here is to ensure condition (\ref{eqn:boundMax}) to be
  met. Possible boundary or singularity problems are circumvented by
  ignoring negligible bits of parameter space via specification
  of~$\epsilon$.  Lipschitz continuity of the divergence will ensure
  that the relevant range may be covered using a finite number of
  bins.  Note that the actual form of the latent (mixing) distribution
  is only used to determine the relevant range in parameter space,
  while the actual binning is otherwise independent.  A number of
  variations of the \textsc{direct} algorithm are conceivable; for
  example, it may or may not be sensible, or possible, to either have
  a reference point or a bin margin at the parameter space's
  boundary. Also, the relationship between $x$ and $p(y|x)$ may not
  necessarily be monotonic, in which case it may be possible to devise
  more efficient non-sequential binning strategies.

\section{Examples}
\subsection{Student-$t$ distribution}\label{sec:StudenttExample}
  A prominent example of a mixture distribution is the
  Student\mbox{-}$t$ distribution.  It arises as a continuous mixture
  of normal distributions with zero mean and scale
  $\sigma=\sqrt{\frac{\nu}{s}}$, where $s$~is a draw from a
  $\chi^2$~distribution with $\nu$~degrees of freedom
  \citep[Ch.~28]{JohnsonKotzBalakrishnan}.  We can approximate the
  marginal Student\mbox{-}$t$ distribution by a mixture of normal
  distributions, conditioning on a finite set of grid points in~$s$,
  and compare against the true marginal which in this case we know to
  be a Student\mbox{-}$t$ distribution.

  \begin{figure}
    \begin{center}
      \includegraphics[width=0.75\linewidth]{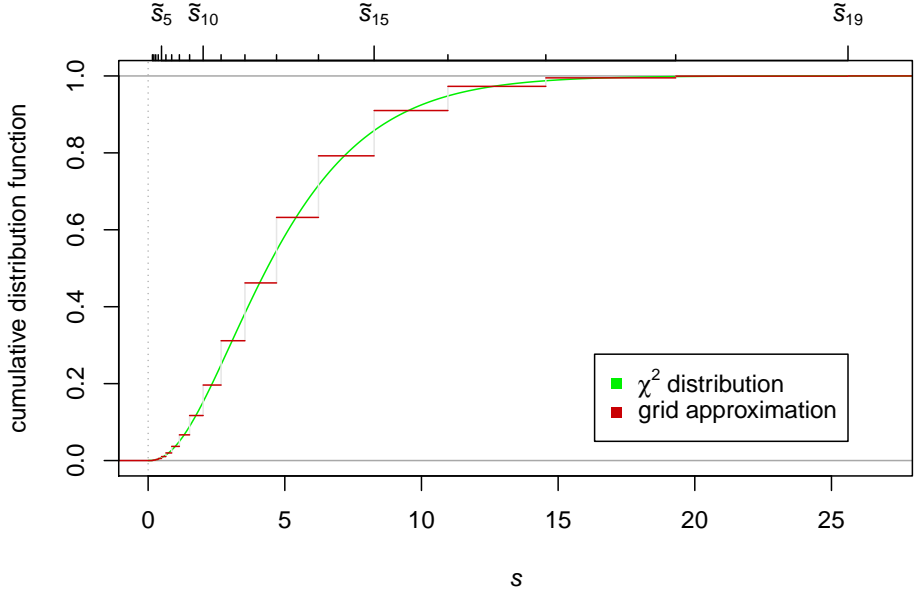} 
      \caption{\label{fig:StudenttGrid}The underlying $\chi^2$
        mixing distribution (of the latent variable~$s$) and the grid
        approximation that is effectively used instead in the
        Student\mbox{-}$t$ example example
        (Sec.~\ref{sec:StudenttExample}). The extra tick marks at the
        top indicate the 19~grid points used.}
    \end{center}
  \end{figure}

  \begin{figure}
    \begin{center}
      \includegraphics[width=0.75\linewidth]{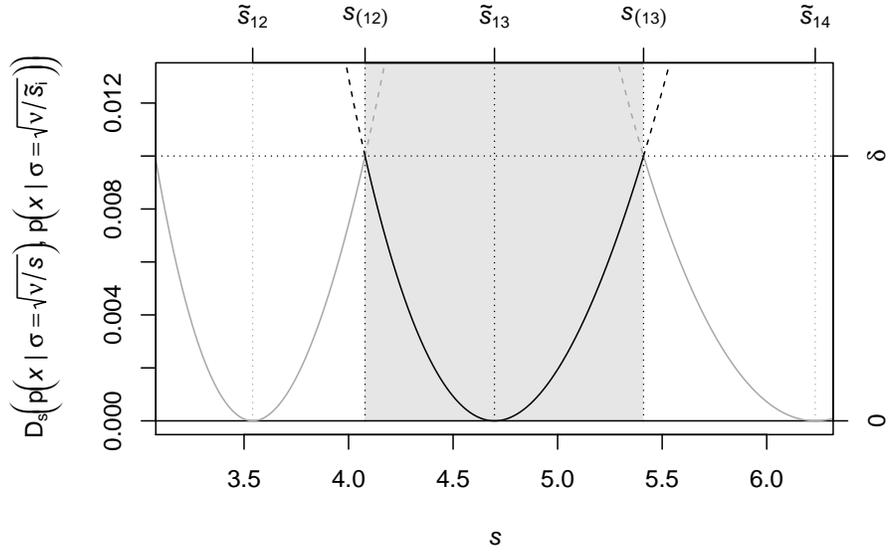} 
      \caption{\label{fig:StudenttBinning}Illustration of how the
        binning is set up (Student\mbox{-}$t$ example,
        Sec.~\ref{sec:StudenttExample}). Bin margins~$s_{(i)}$ and reference
        points~$\tilde{s}_i$ are arranged such that within each bin
        the divergence relative to the corresponding reference point
        does not exceed the pre-set threshold~$\delta$.}
    \end{center}
  \end{figure}

  \begin{figure}
    \begin{center}
      \includegraphics[width=0.9\linewidth]{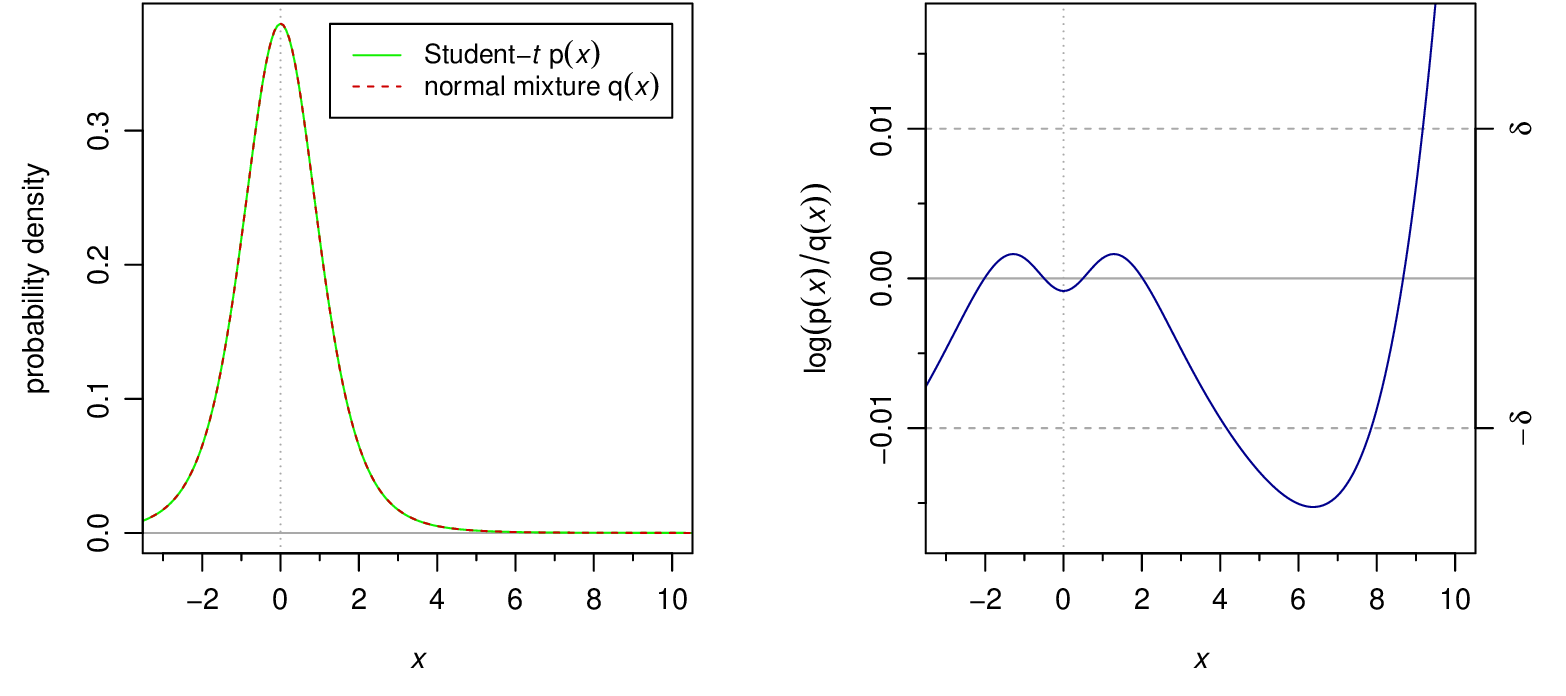} 
      \caption{\label{fig:StudenttApprox}Comparison of the true
        mixture distribution, the Student\mbox{-}$t$
        distribution, to the grid approximation. The left
        panel shows the two probability density functions on top of
        each other; the two are essentially undiscernible at this
        scale. The right panel shows the logarithmic ratio of the
        densities as a function of~$x$.}
    \end{center}
  \end{figure}

  Suppose we are interested in the case of $\nu=5$~degrees of freedom.
  We set the tuning parameters to $\delta:=0.01$ and $\epsilon:=0.001$
  and we use the $\chi^2_5$ distribution's
  $\frac{\epsilon}{2}$-quantile as the starting reference point
  ($\tilde{s}_1:=0.158$).  Applying the sequential DIRECT algorithm
  from Sec.~\ref{sec:algorithm} (utilizing
  expression~(\ref{eqn:normalDiv})) results in a set of 19 reference
  points~$\tilde{s}_i$.
  As a result from the implied differences in the corresponding
  conditionally normal distributions, the 19~reference points are very
  unequally spaced, with many points concentrated near zero and a
  coarser spacing at large values (see Fig.~\ref{fig:StudenttGrid}).

  Fig.~\ref{fig:StudenttBinning} illustrates the construction of the
  binning by showing the 13th bin and its two neighbouring bins with
  bin margins~$s_{(i)}$ and reference points~$\tilde{s}_i$.  One can
  see that by construction within each bin the divergence relative to
  the corresponding reference point,
  $\symKL\bigl(p(x|\sigma\!=\!\sqrt{\nu/s}) \big\|
  p(x|\sigma\!=\!\sqrt{\nu/\tilde{s}_i})\bigr)$, 
  remains below~$\delta$.

  The 19-component normal mixture approximation is compared to the
  true marginal distribution in Fig.~\ref{fig:StudenttApprox}.  The
  two densities are barely distinguishable, and their ratio is very
  close to unity; it only diverges towards the distributions' extreme
  tails.  The numerically computed actual divergence in this case
  amounts to 
  $\symKL\bigl(p(x) \big\| q(x)\bigr) \approx 3.5 \!\times\! 10^{-5}$.

\subsection{Convolution of two distributions}\label{sec:ConvExample}
  In the following we present the example of computing the convolution
  of two distributions.  Suppose we have two random variables, $X$ and
  $Y$, with densities $p_X(x)$ and $p_Y(y)$. We are interested in
  their sum $Z=X+Y$, and its density $p_Z(z)$.
  Here we take $X$ and $Y$ to follow skew-normal and logistic
  distributions, respectively, so that the solution is not trivial.
  We can turn the problem into that of a mixture distribution and
  subsequently apply the above algorithm by first considering the
  joint distribution of $X$ and $Z$.  Note that $\prob(Z=z\,|\,X=x)
  \;=\; \prob(Y=z-x)$, so the conditional distribution of $Z|X$ here
  is simply a ``shifted'' version of the (known) distribution
  $\prob_Y$.  With that, we can rewrite the target density $p_Z$ as a
  marginal density in terms of the (known) marginal $p_X(x)$ and the
  (known) conditional $p_Z(z|x) = p_Y(z-x)$:
  \begin{equation}
    p_Z(z) \;=\; \int  p_Y(z-x) \,p_X(x) \,\differential x \mbox{.}
  \end{equation}
  This way it is obvious that convolution of two random variables may
  again be seen as a special case of a mixture distribution where the
  conditional $\prob(Z|X)$ is mixed via the latent distribition
  $\prob(X)$. Due to symmetry of the problem, the roles of $X$ and $Y$
  may also be reversed.

  In the following suppose that $p_X(x)>0$ and $p_Y(y)>0$ for all
  $x,y\in\realline$, i.e., the domain of both $X$ and $Y$ is the whole
  real line.
  \begin{figure}[t]
    \begin{center}
      \includegraphics[width=0.6\linewidth]{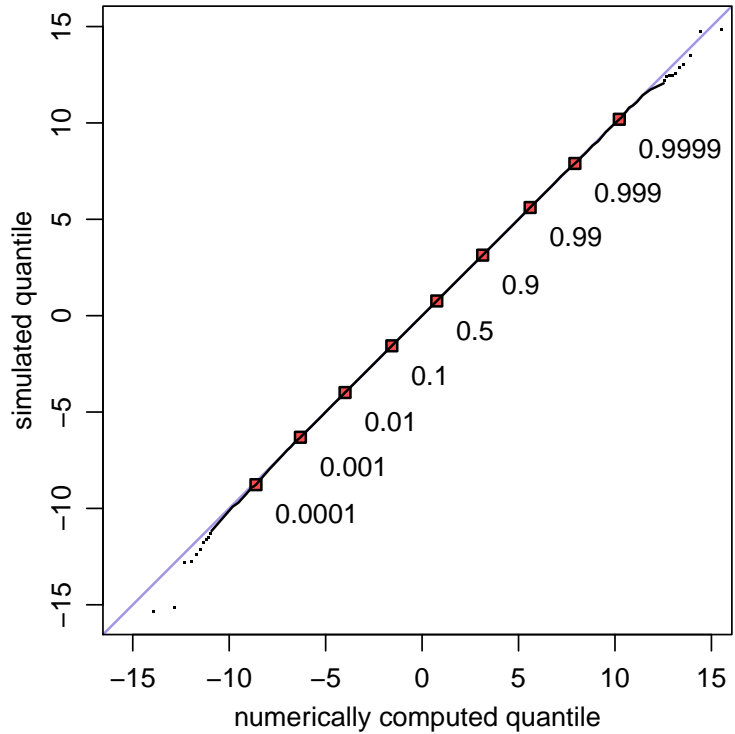} 
      \caption{\label{fig:qqplot}Q-Q-plot illustrating the accuracy of
        the convolution of a skew-normal and a logistic distribution
        by comparing quantiles computed numerically, using the
        \textsc{direct} algorithm, and simulated quantiles.}
    \end{center}
  \end{figure}
  When applying the \textsc{direct} algorithm to set up an
  approximation, it is important to note that the divergence $\symKL
  \bigl( p_Z(z|x_1) \big\| p_Z(z|x_2)\bigr)$
  required in steps 2--4 of the algorithm (Sec.~\ref{sec:algorithm})
  only depends on the (absolute) difference $|x_2\!-\!x_1|$,
  since the conditional distributions $p_Z(z|\,\cdot\,)$ here only
  differ by a shift in location.
  This implies that the bin width $\tilde{x}_{i} - \tilde{x}_{i-1}$ is
  constant across all bins, and hence only needs to be determined
  once. This simplifies the grid construction to a few steps:
  \begin{enumerate}
    \item determine the bin half-width 
          $\Delta_x$ such that 
          $\symKL \bigl( p_Y(y) \big\| p_Y(y-\Delta_x)\bigr) = \delta$.
    \item determine minimum and maximum $X$ 
          values $\tilde{x}_1$ and $\tilde{x}_k$
          e.g. as the $\frac{\epsilon}{2}$ 
          and  $1\!-\!\frac{\epsilon}{2}$ quantiles of $p_X$.
    \item determine the remaining reference points $\tilde{x}_2$ to
      $\tilde{x}_{k-1}$ as well as their total number~$k$ by filling the
      interval with reference points that are at most 
      $(\tilde{x}_{i} - \tilde{x}_{i-1}) \leq \Delta_x$ apart.
  \end{enumerate}
  A general implementation of the procedure in~\textsf{R} is shown in
  the online supplement.  Divergences here are computed numerically,
  without needing to have the corresponding formulas available in
  analytic form.

  Consider the example of the sum of two random variables, one
  following a skew-normal distribution with shape parameter $\alpha=4$
  \citep{Azzalini,AzzaliniWebsite}, and one following a logistic
  distribution.  Application of the \textsc{direct} algorithm (using
  $\delta=0.01$ and $\epsilon=0.001$) results in a 13-component
  mixture of logistic distributions to approximate the convolution.
  Draws from the two summands' distributions may easily be simulated,
  so it is straightforward to also generate samples of their sum's
  distribution.  Figure~\ref{fig:qqplot} illustrates the fit of the
  numerical approximation to $1\,000\,000$ simulated samples via a
  quantile-quantile plot (Q-Q plot).  Here the 10 smallest and largest
  samples are shown as individual dots, other quantiles are connected
  by a line, and selected quantiles are highlighted.  Note that while
  the design parameter $\epsilon$ was set to $0.001$, the simulated
  and computed quantiles appear to match well even beyond tail
  probabilities of $0.001$.  The \textsf{R}~code to reproduce these
  simulations is also provided in the online supplement.  All
  computations here were carried out using~\textsf{R}
  \citep{RManual2015}.

\section{Conclusions}\label{sec:conclusions}
  The \textsc{direct} approach introduced in this paper allows to
  generate finite mixtures as approximations to mixture distributions
  with a large or infinite number of mixture components.  A
  formulation in terms of a finite mixture distribution then makes
  density function, cumulative distribution function, etc.\ easily
  accessible.  The mismatch incurred by resorting to the approximation
  is efficiently controlled via two tuning parameters ($\delta$
  and~$\epsilon$). The described algorithm allows for easy
  implementation in a completely automated fashion, as is also
  demonstrated in the examples.  The setup relies on the computation
  of (symmetrized) divergences of (conditional) distributions; ideally
  these are available analytically, but \mbox{numerical} computation
  is also not a problem.

  Variations of the \textsc{direct} algorithm are conceivable.  The
  bound derived in Sec.~\ref{sec:construct}
  may be met in many different ways; the described one is only a
  simple, general solution. For example, it may be possible, and
  possibly more efficient, to aim at the condition in
  (\ref{eqn:boundSum}) rather than (\ref{eqn:boundMax}) in order to
  bound the divergence. While for simplicity we concentrated on
  symmetrized divergences here, it may also make sense to directly aim
  for directed Kullback-Leibler divergences instead.

  A generalization to higher dimensions of the latent mixing
  distribution should in general also be possible. Since the problem
  of covering of higher-dimensional spaces is considerably trickier,
  it may eventually be easiest to resort to random coverings here
  \citep{MessengerPrixPapa2009,Roever2010}.

  The algorithm was originally developed and eventually applied in the
  context of the \texttt{bayesmeta} \textsf{R} package
  \citep{bayesmeta}.  In this meta-analysis application, one is faced
  with the common problem of inferring two parameters ($\tau$
  and~$\mu$) via their posterior probability distribution.  From their
  joint distribution ($p(\mu,\tau)$) one of the marginals, $p(\tau)$,
  may be derived analytically, while the conditionals $p(\mu|\tau)$
  are normal.  Primary interest usually lies in $\mu$, and application
  of the \textsc{direct} algorithm facilitates quick and accurate
  computation of the marginal $p(\mu)$ without having to use, for
  example, Markov chain Monte Carlo (MCMC) methods
  \citep{SimulationPaperDummyRef}.

{
  \bibliographystyle{agsm}
  \bibliography{mixture}
}

\end{document}